\documentclass[a4paper,fleqn]{cas-sc}

\usepackage[numbers]{natbib}
\usepackage{siunitx}
\usepackage{hyperref}
\usepackage{cleveref}
\usepackage{lineno}

\def\tsc#1{\csdef{#1}{\textsc{\lowercase{#1}}\xspace}}
\tsc{WGM}
\tsc{QE}

\begin{document}
\let\WriteBookmarks\relax
\def\floatpagepagefraction{1}
\def\textpagefraction{.001}

\shorttitle{The RD50-MPW4 HV CMOS Sensor}    

\shortauthors{Jory Sonneveld et. al}  

\title [mode = title]{The RD50-MPW4: A Radiation Hard HV CMOS Sensor for Future Colliders}  

\tnotemark[1] 

\tnotetext[1]{} 

%

\author[1]{Jory Sonneveld}

\cormark[1]


\ead{jory.sonneveld@nikhef.nl}



\affiliation[1]{organization={Nikhef National Institute for Subatomic Physics},
            addressline={Science Park 105}, 
            city={Amsterdam},
            postcode={1098XG}, 
            country={The Netherlands}}
\author[2]{Thomas Bergauer}





\affiliation[2]{organization={MBI - Marietta Blau Institute for Particle Physics
of the Austrian Academy of Sciences},
            addressline={Dominikanerbastei 16, 3rd floor}, 
            city={Vienna},
            postcode={1010}, 
            country={Austria}}



\author[3]{Raimon Casanova}
\affiliation[3]{organization={Institute for High Energy Physics (IFAE), Autonomous University of Barcelona (UAB)},
            addressline={Bellaterra}, 
            city={Barcelona},
            postcode={08193}, 
            country={Spain}}
\author[2]{Harald Handerkas}
\author[2]{Christian Irmler}
\author[4]{Jorge Jiménez-Sánchez}
\author[1]{Uwe Krämer}
\fnmark[1] 
\fntext[1]{Now at TÜ Dortmund.}
\author[5]{Ricardo Marco-Hernandez}
\affiliation[5]{organization={Instituto de Física Corpuscular (IFIC), CSIC-UV.},
            addressline={Parque Científico, Catedrático José Beltrán, 2}, 
            city={ Paterna (Valencia)},
            postcode={46980}, 
            country={Spain}}
\author[5]{José Mazorra de Cos}
\author[4]{Fernando Muñoz-Chavero}
\author[4]{Rogelio Palomo}

\affiliation[4]{organization={Dept. of Electronic Engineering, School of Engineering at University of Sevilla},
            addressline={Avda. de los Descrubrimientos s/n}, 
            city={Sevilla},
            postcode={41092}, 
            country={Spain}}
\author[2]{Bernhard Pilsl}
\author[2]{Sebastian Portschy}
\author[6]{Samuel Powell}
\affiliation[6]{organization={Department of Physics, University of Liverpool, Oliver Lodge Building},
            addressline={Oxford Street}, 
            city={Liverpool},
            postcode={L69 7ZE}, 
            country={UK}}
\author[2]{Patrick Sieberer}
\fnmark[2] 
\fntext[2]{Now at Paul Scherrer Institut, Villigen.}
\author[2]{Helmut Steininger}
\author[6]{Eva Vilella}
\author[6]{Benjamin Wade}
\author[6]{Chenfan Zhang}
\author[7]{Sinuo Zhang}
\affiliation[7]{organization={Physikalisches Institut, Rheinische Friedrich-Wilhelms-Universitaet Bonn},
            addressline={Nussallee 12}, 
            city={Bonn},
            postcode={53115}, 
            country={Germany}}

\begin{abstract}
The former CERN RD50 collaboration develops monolithic active pixel high voltage (HV) CMOS sensors for future colliders with the aim of high radiation tolerance, good time resolution, and high granularity pixel detectors. The most recent prototype, the RD50-MPW4, was produced by LFoundry in December 2023 using a 150 nm CMOS process. It features a matrix of 64x64 pixels with a \SI{62}{\micro\meter} pitch and employs a column-drain readout architecture. Compared to its predecessor, it now has separate analog and digital power domains and a new biasing scheme with a guard ring structure that supports bias voltages over 600~V.

This contribution will discuss the design and latest results of the MPW4, where tests with unirradiated samples showed more than 99.9\% efficiency, \SI{16}{\micro\meter} spatial resolution and 10~ns timing resolution. Efficiencies of over 99\% were achieved for samples irradiated to fluences of $1 \times 10^{15}$~1~MeV n$_{\mathrm{eq}}/\mathrm{cm}^2$. A 3D map of the charge collection efficiency from a measurement with two-photon absorption laser is presented that nicely outlines the depletion depth of this radiation hard, high granularity monolithic active pixel sensor.
\end{abstract}




\maketitle

\section{Radiation Hard HV CMOS Monolithic Sensors for Particle Tracking}
\label{sec:introduction}
The former RD50 Collaboration developed radiation hard semiconductor devices for very high luminosity colliders. The CMOS working group in this collaboration developed four prototypes of depleted monolithic active pixel sensors (DMAPS) on multiproject wafers (MPWs) in the LFoundry S.r.l. 150 nm CMOS technology \cite{rd50hvcmos} 
 for use in tracking and vertexing in future high energy physics experiments with the aim of excellent radiation tolerance, fast timing resolution \cite{mpw2timeres} and high granularity. These DMAPS, which also called high voltage (HV) CMOS devices, are thin sensors with electronics integrated into the sensor, and a high voltage applied for a large depletion region and high radiation tolerance. Since the third in the series, the RD50-MPW3, the prototype sensors feature large 64x64 pixel matrices and a low voltage differential signal (LVDS) digital readout at 640 Mb/s. Since the last prototype, the RD50-MPW4, the limit of the breakdown voltage has been pushed significantly higher with the implementation of advanced ring structures based on the design of the LF-Monopix2 prototype \cite{dmaps180} as was predicted by \cite{guardrings}. With an operating voltage of more than 600~V this has pushed the radiation tolerance of these devices to much higher levels.
 \begin{figure}
    \centering
    \includegraphics[width=0.47\linewidth]{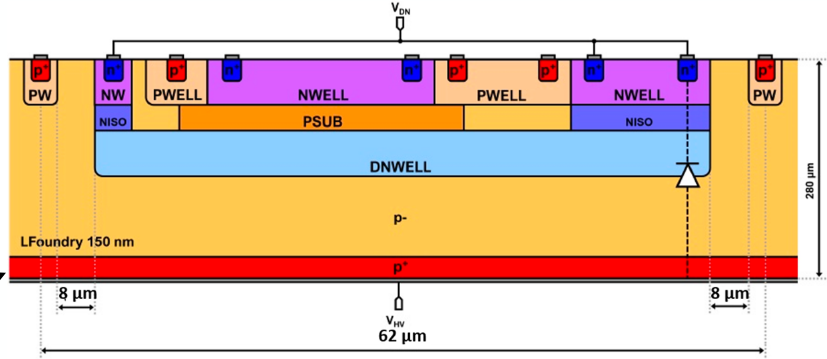}
    \includegraphics[width=0.17\linewidth]{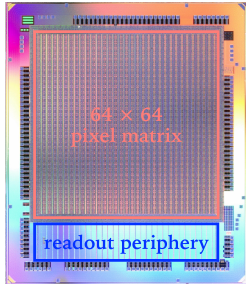}\hspace{0.02\linewidth} 
    \includegraphics[width=0.32\linewidth]{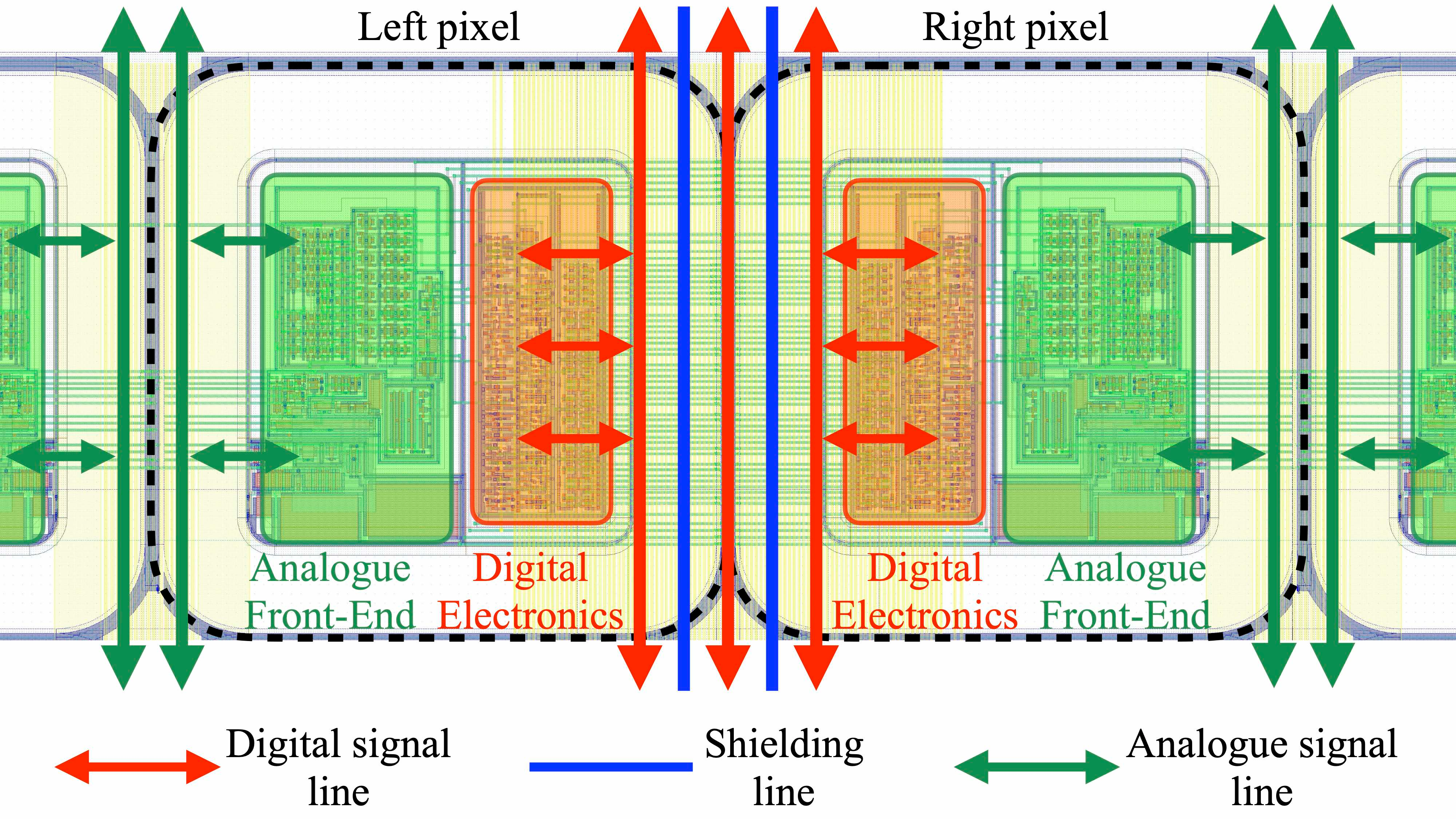}
    \caption{Left: The RD50-MPW4 pixel layout for postprocessed sensors with backside biasing. This large collection electrode sensor has a deep n-well (DNWELL) in a p-substrate. A deep p-well (PSUB) for shielding enables CMOS electronics to be placed in the pixel. Figure from \cite{mpw4}. Middle: Layout of the RD50-MPW4. The sensor features a 64 by 64 pixel matrix and a readout periphery. Right: Signal lines for two pixels in a double column. Figures from \cite{mpw4presentationchenfan}.}
    \label{fig:mpw4pixel}
\end{figure}

\section{The RD50-MPW4 sensor and Leakage Current}
\label{sec:rd50mpw4}
The RD50-MPW series are large collection electrode monolithic active pixel sensors with a deep n-well in a p-substrate, with the readout electronics embedded inside the deep n-well, as shown in \cref{fig:mpw4pixel}. A deep p-well shields the collection electrode so that CMOS electronics can be placed inside the pixel. For the RD50-MPW4, wafers have a nominal resistivity of $3~$k$\Omega\cdot$cm and are thinned to \SI{280}{\micro\meter}. The substrate can be biased with a high voltage of over 600 V to improve tolerance to radiation damage from non-ionizing energy loss (NIEL). This high voltage is applied from the top side. As was shown in \cite{cmosthesis}, backside-biased HV-CMOS sensors have a stronger, more uniform electric field which improves the radiation tolerance \cite{etcthvcmos,ccecmos}. For this reason, two of the three were backside processed to enable backside biasing. 

\begin{figure}
    \centering
    \includegraphics[width=0.3\linewidth]{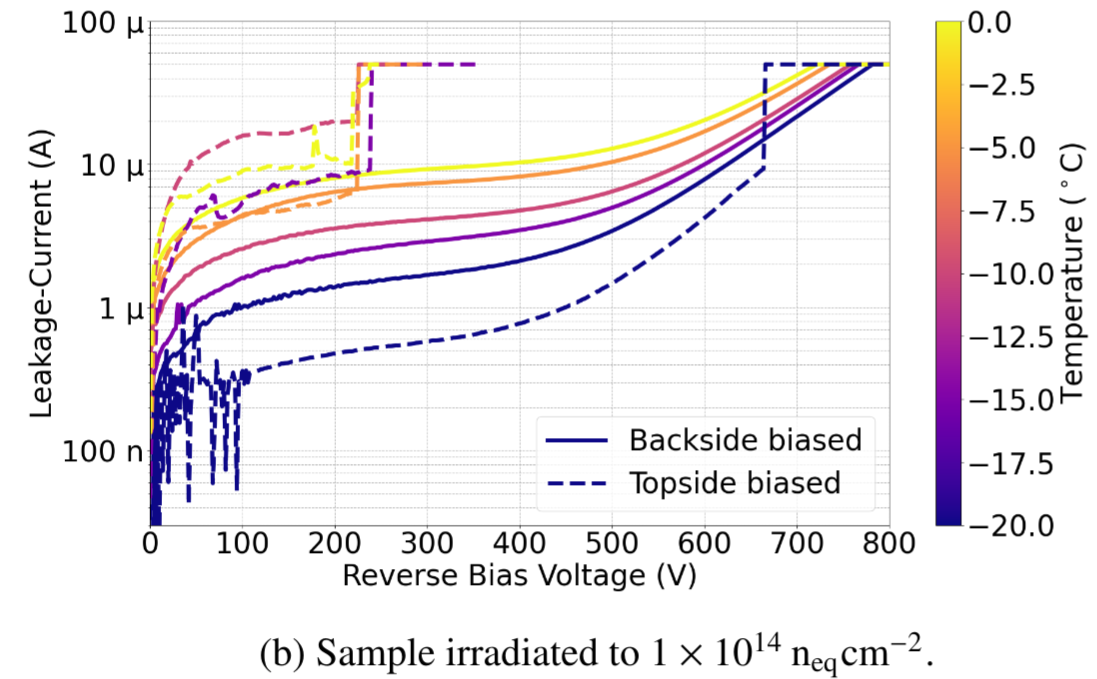}
        \includegraphics[width=0.3\linewidth]{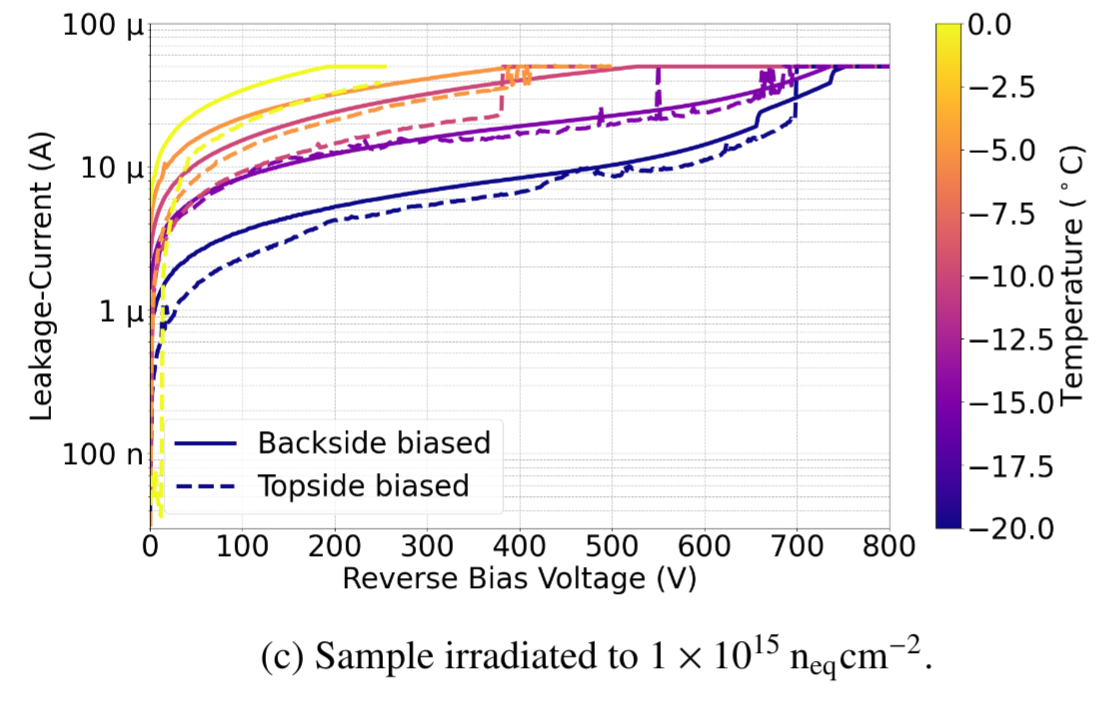}
    \includegraphics[width=0.3\linewidth]{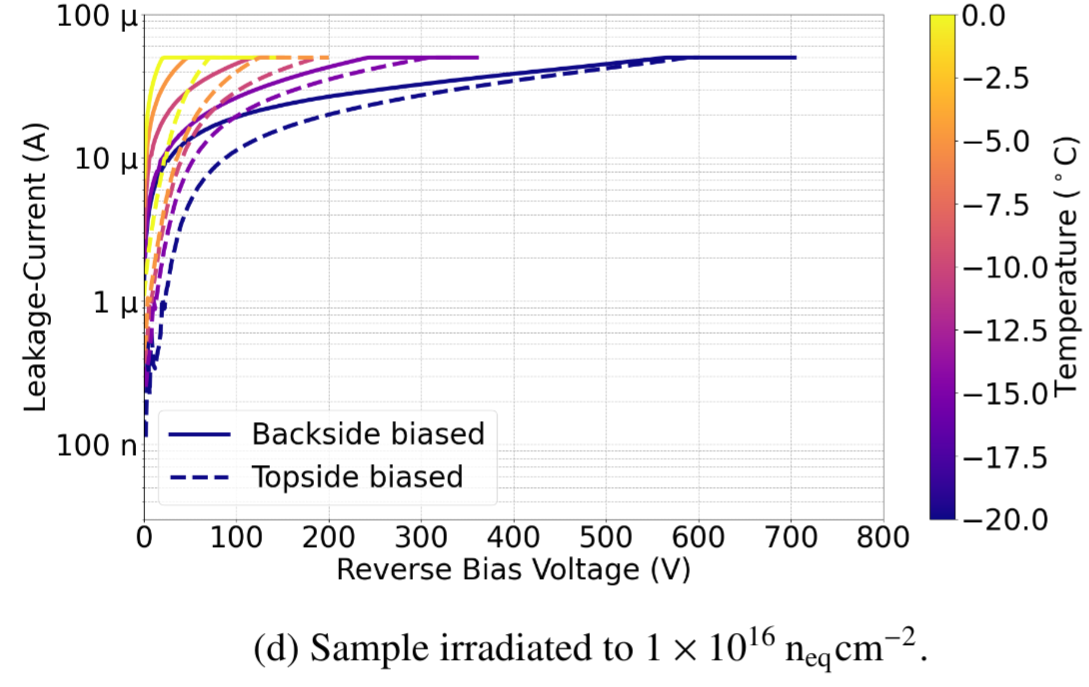}
    \caption{Leakage current in RD50-MPW4 sensors irradiated to fluences of $\Phi = 1 \times 10^{14}~\mathrm{n}_{\mathrm{eq}}/\mathrm{cm}^2$ (left), $\Phi = 1 \times 10^{15}~\mathrm{n}_{\mathrm{eq}}/\mathrm{cm}^2$ (middle),$\Phi = 1 \times 10^{16}~\mathrm{n}_{\mathrm{eq}}/\mathrm{cm}^2$ (right) for different temperatures. Sensors show increased leakage current for increased temperatures as expected, and can be biased beyond 600 V. Figures from \cite{vciproceedings}.}
         \vspace{-0.5cm}
    \label{fig:ileak}
\end{figure}

The leakage current of a sensor gives an indication its radiation tolerance. Higher leakage current results in more noise and less radiation tolerance. For the RD50-MPW4 it was shown that unirradiated samples have a leakage current of the order of 10~nA at a depletion voltage of -190~V; see \cref{fig:ileak}. The sensors have been shown to be operated safely at biases beyond 600~V, and the leakage current increased for increased temperatures as expected \cite{vciproceedings}.

\section{The RD50-MPW4 Electronics and Readout}
The RD50-MPW4 sensors have $64\times64$ pixel matrices and a periphery for the digital readout, as shown in the middle in \cref{fig:mpw4pixel}. Pixels are of \SI{62}{\micro\meter} pitch and feature a digital and analog readout inside the pixel. The pixel matrix has a double-column architecture. A double column is lined by analog signal lines on the side and digital signal lines and shielding lines in the middle to reduce cross talk between the lines \cite{mpwseries}, as shown on the right in \cref{fig:mpw4pixel}.

\begin{figure}
    \centering
    \includegraphics[width=0.8\linewidth]{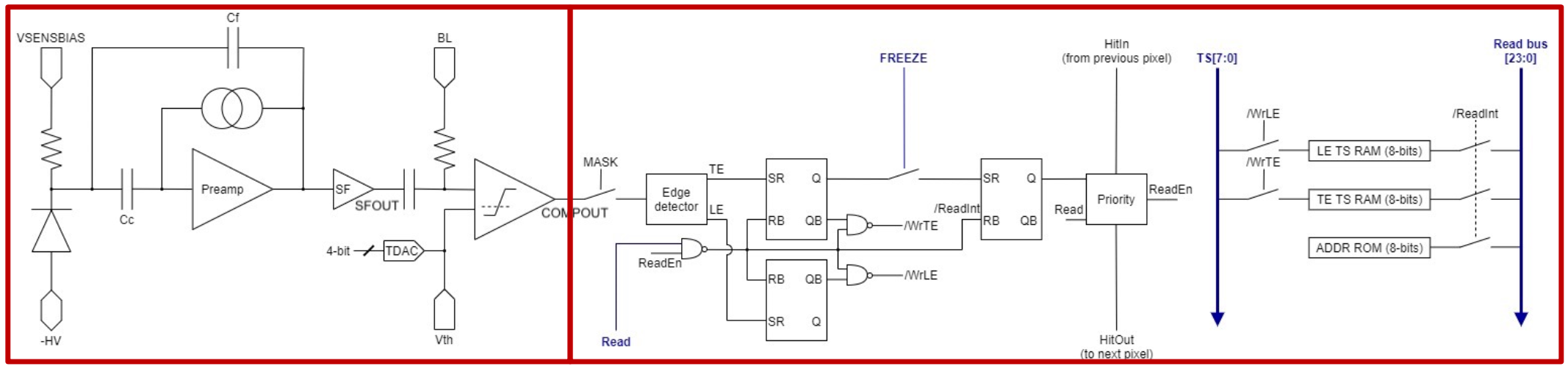}
    \caption{In-pixel electronics of the RD50-MPW4. The pixel features analog readout up to the comparator output (left box), after which a fully digital signal is routed out of the pixel towards the chip periphery (right box). Figure from 
    \vspace{-0.5cm}
    \cite{mpw4presentationchenfan}.}
    \label{fig:pixelelectronics}
\end{figure}

\begin{figure}
    \centering
    \includegraphics[width=0.4\linewidth]{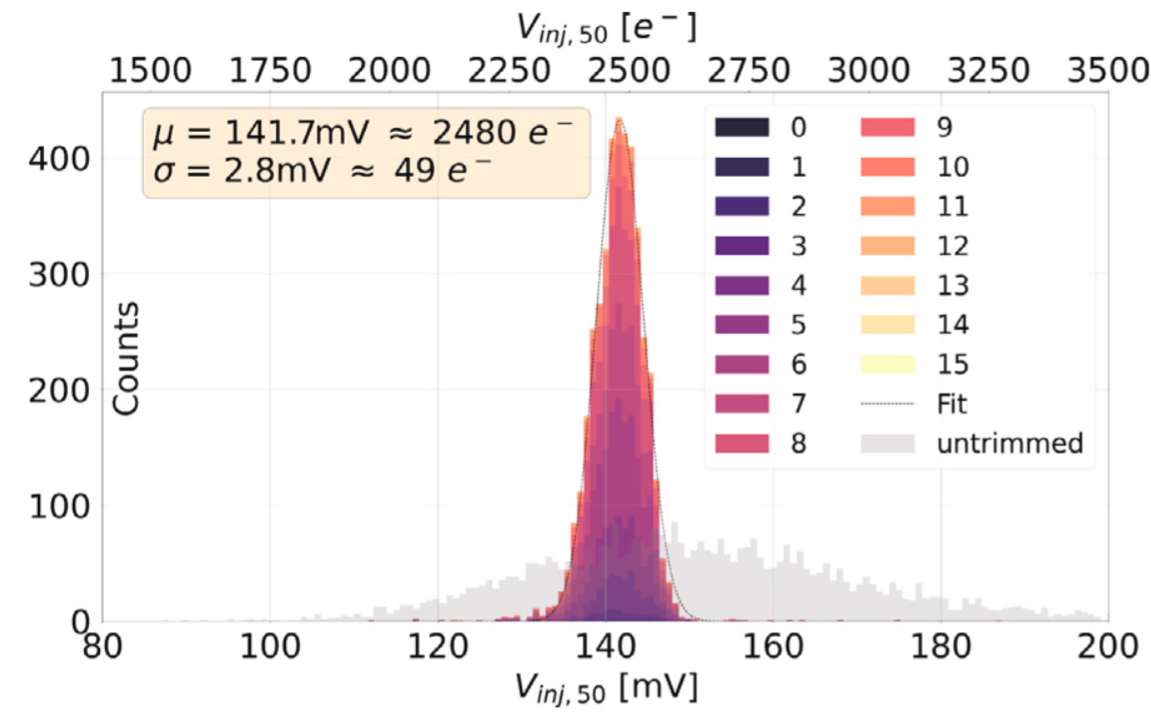}
    \includegraphics[width=0.4\linewidth]{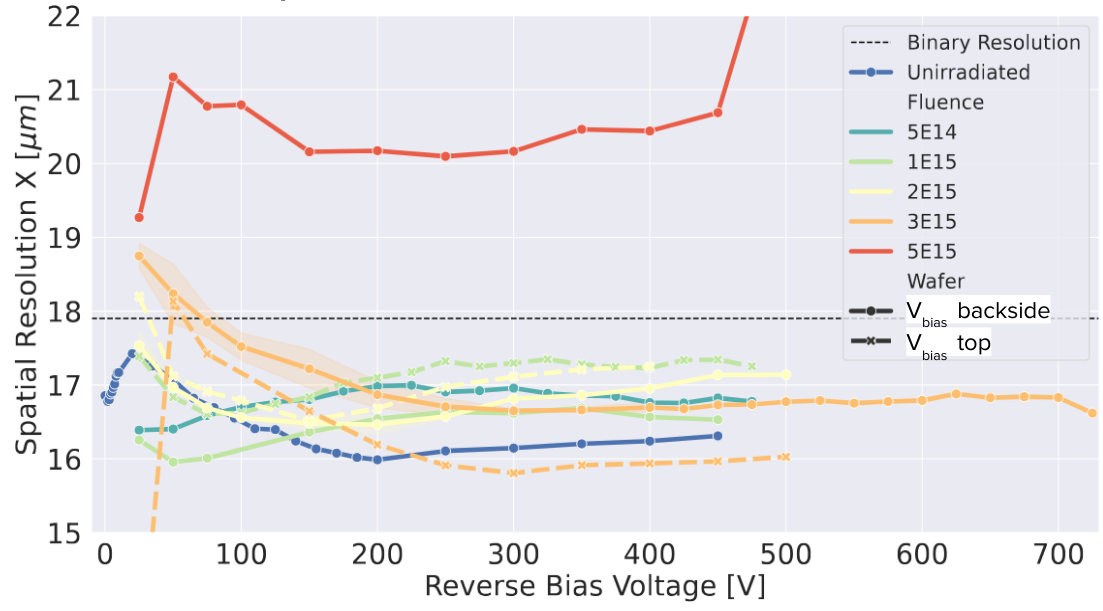}
    \caption{Left: Thresholds of an RD50-MPW4 before and after trimming \cite{mpw4characterization}. Right: Spatial resolution for sensors irradiated to different fluences. All except the most highly irradiated have a better than binary resolution.}
    \label{fig:res}
\end{figure}

The RD50-MPW4 in-pixel electronics layout is shown in figure \cref{fig:pixelelectronics}. It features a fully digital readout and an in-pixel charge sensitive amplifier with injection circuit. A comparator with a 4-bit trim digital to analog converter (DAC) can be used to trim individual pixel thresholds relative to the value that is set centrally for the entire chip. An example of a chip with trimmed thresholds is shown on the left in \cref{fig:res}. From simulations, the processing time for a minimum-ionizing-particle-like signal of 25 ke$^-$ is less than 200 ns. The chip features a triggerless column drain readout architecture similar to that of the ATLAS FEI3. The row address of a pixel is sent out to the periphery where the end-of-column information is appended to form an 8-bit pixel address. This then contains the row (pixel) and column address. Further 8-bit leading and trailing edges give the time over thresholds of the signal.

The RD50-MPW4 has an I2C protocol for slow control configuration. Each of the 32 end-of-columns (EoCs) serialize the data which is then sent out through a 640 Mbit/s low voltage differential signal (LVDS) link. Data acquisition is based on the Caribou system \cite{caribou} that is connected to a FPGA Mezzanine Card (FMC) and a Xilinx ZC706 or ZC702 evaluation board.

\section{Efficiency and Spatial Resolution}
\label{sec:eff}
At a test beam at DESY with 4.2 GeV electrons, 10 sensors were tested with the Adenium telescope \cite{adenium} and a Telepix2 chip \cite{telepix} for trigger and region of interest. Data was acquired using the EUDAQ2 framework \cite{eudaq2} and analyzed using Corryvreckan \cite{corry}. The geometric mean of the biased (including the device under test, or DUT) and unbiased resolution (excluding the DUT) was used \cite{resolution}. Sensors were annealed for 80 minutes at 60 degrees Celsius. 
All samples except for one irradiated to  $\Phi = 5 \times 10^{15}~\mathrm{n}_{\mathrm{eq}}/\mathrm{cm}^2$ have a better than binary resolution, as is shown on the right in \cref{fig:res}.  There is no clear correlation between fluence and resolution, and backside biased and topside biased sensors seem to perform equally well.

\begin{figure}
    \centering
    \includegraphics[width=0.3\linewidth]{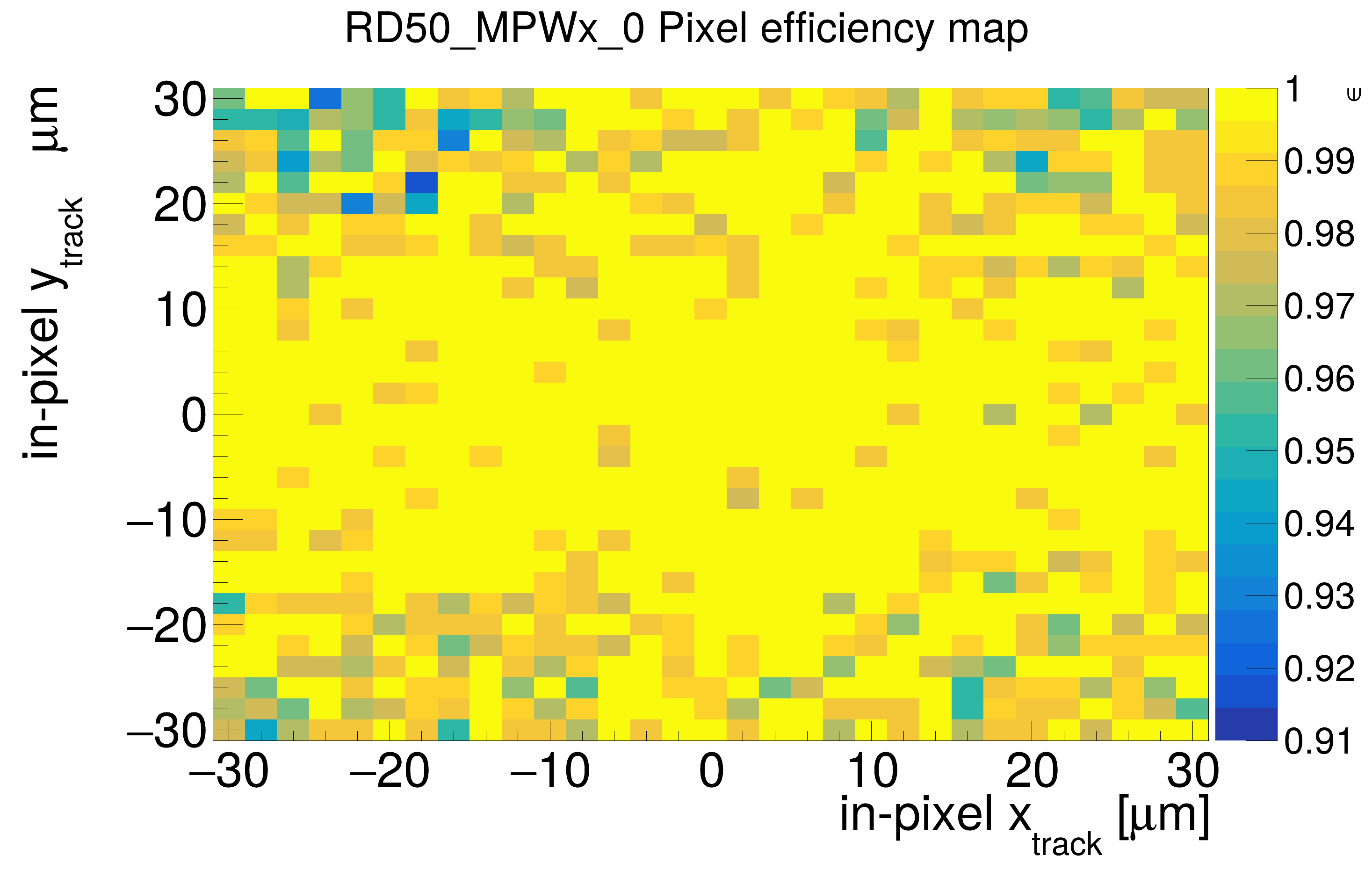}
        \includegraphics[width=0.3\linewidth]{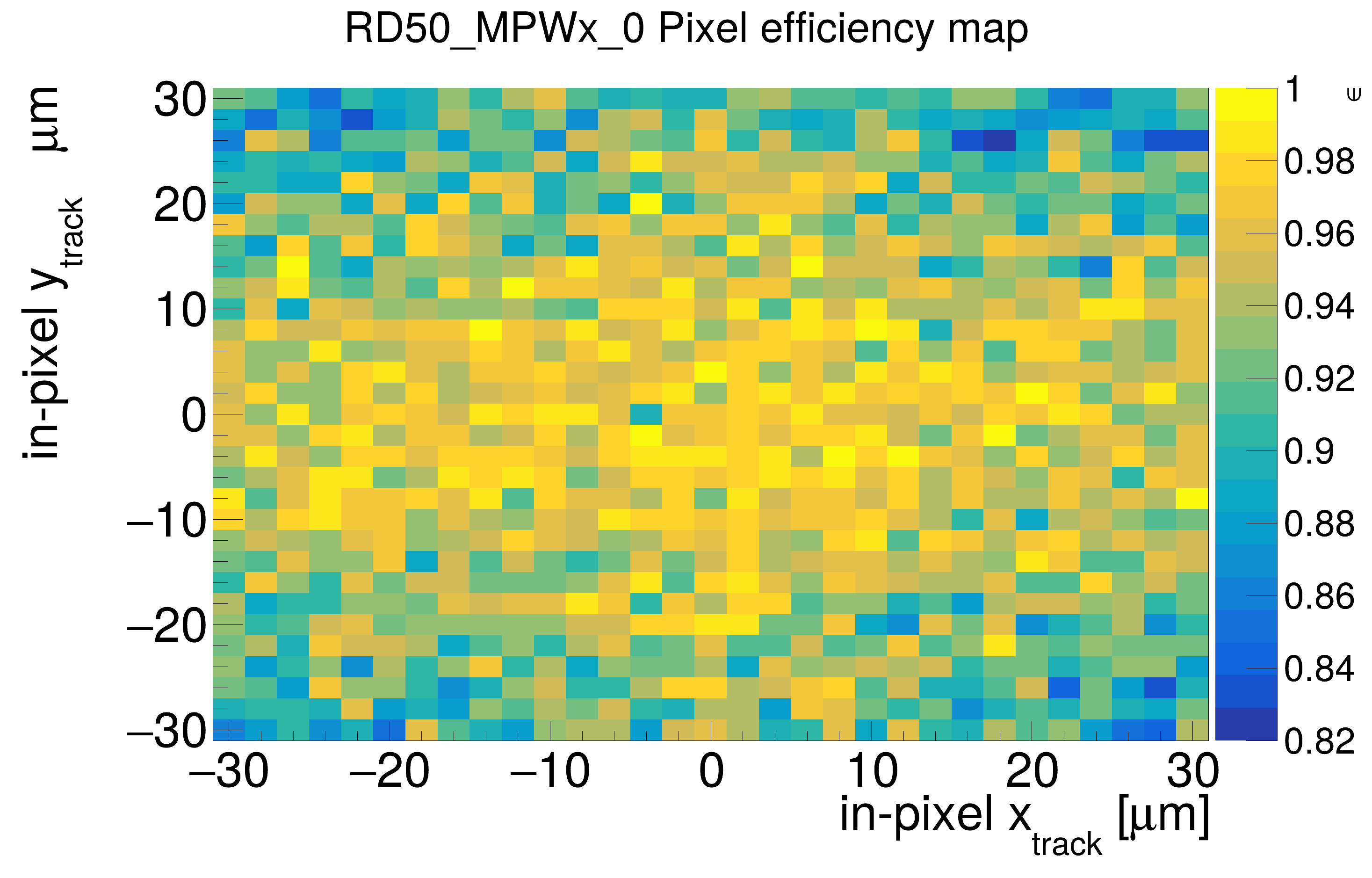}
    \includegraphics[width=0.3\linewidth]{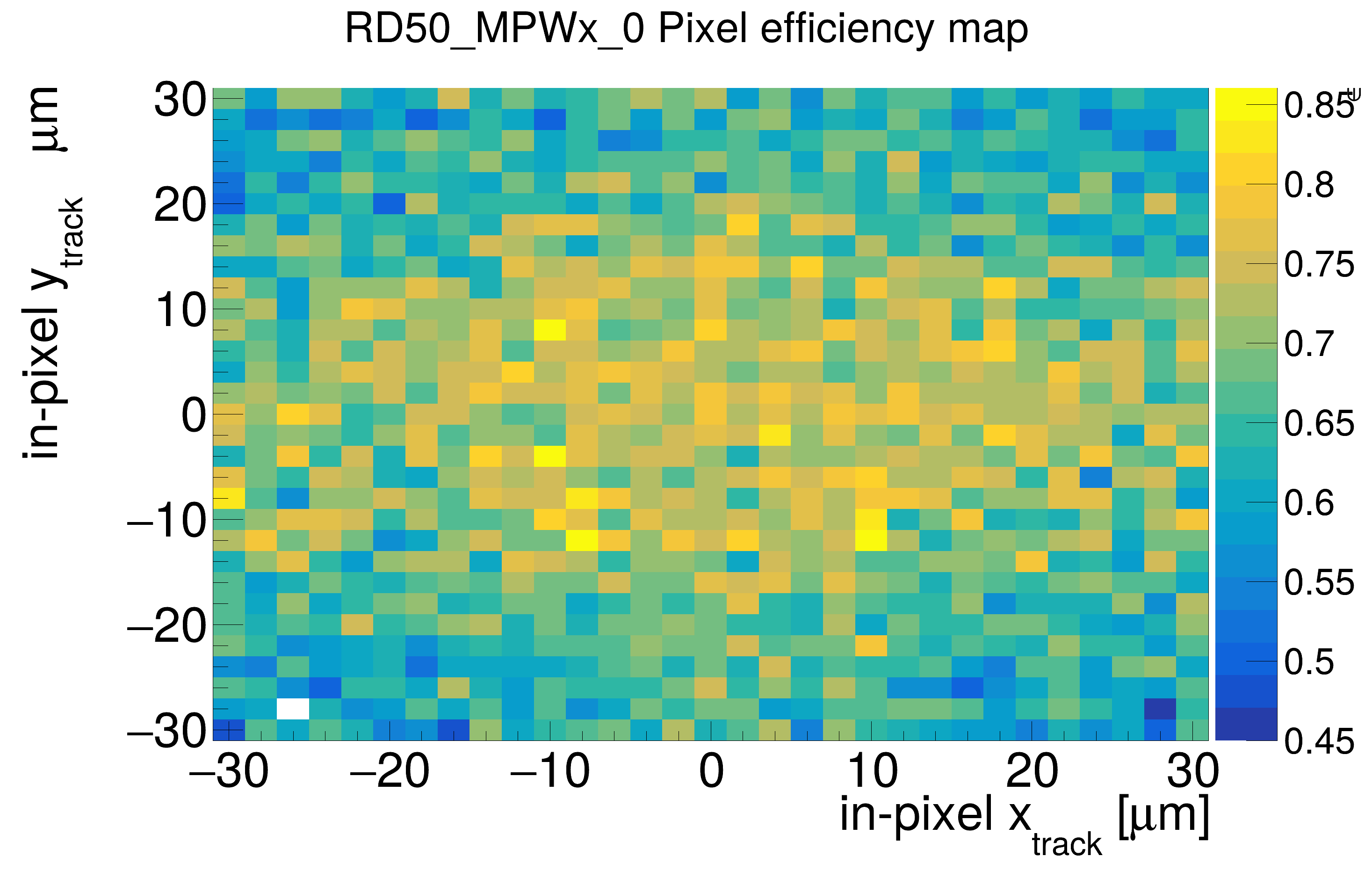}
    \caption{In-pixel efficiency for all pixels of one sensor overlaid for sensors irradatiated to three different fluences: $\Phi = 1 \times 10^{15}~\mathrm{n}_{\mathrm{eq}}/\mathrm{cm}^2$ (left), $\Phi = 2 \times 10^{15}~\mathrm{n}_{\mathrm{eq}}/\mathrm{cm}^2$ (middle), $\Phi = 3 \times 10^{15}~\mathrm{n}_{\mathrm{eq}}/\mathrm{cm}^2$ (right). Efficiency degrades with higher fluence as expected, especially in corners where charge sharing is increased. Figures from \cite{vciproceedings}.}
     \vspace{-0.5cm}
    \label{fig:eff_inpixel}
\end{figure}
The efficiency of the RD50-MPW4 was measured for various fluences. As expected, efficiency degrades with higher fluences, especially in the pixel corners as a result of charge sharing. This in-pixel efficiency, of all pixels of the matrix overlaid, is shown in \cref{fig:eff_inpixel}.

\begin{figure}
    \centering
    \includegraphics[width=0.4\linewidth]{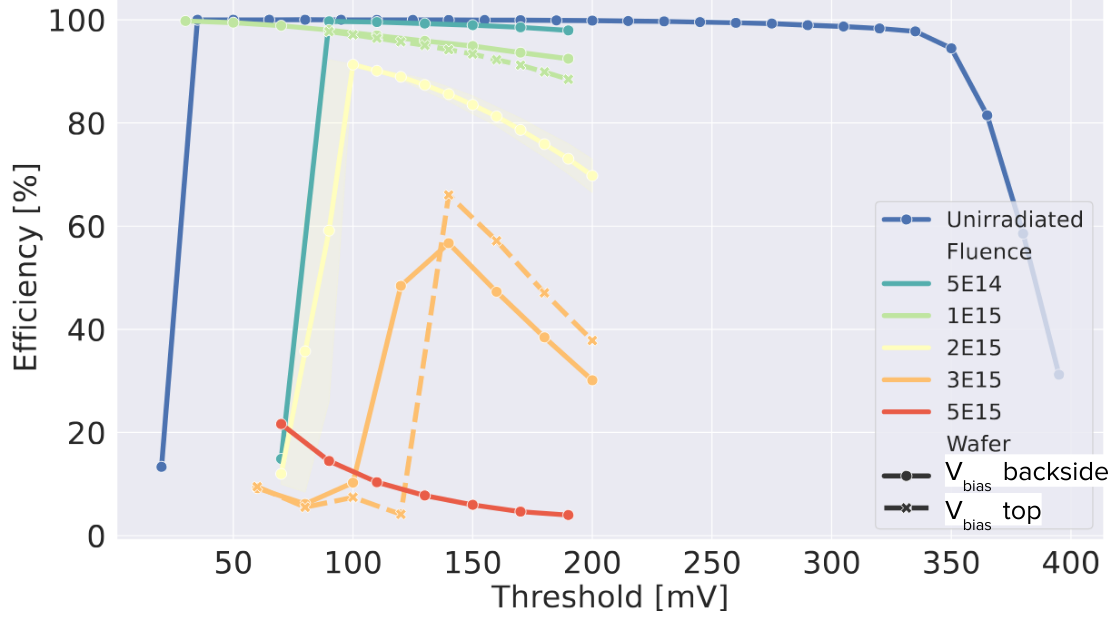}
        \includegraphics[width=0.4\linewidth]{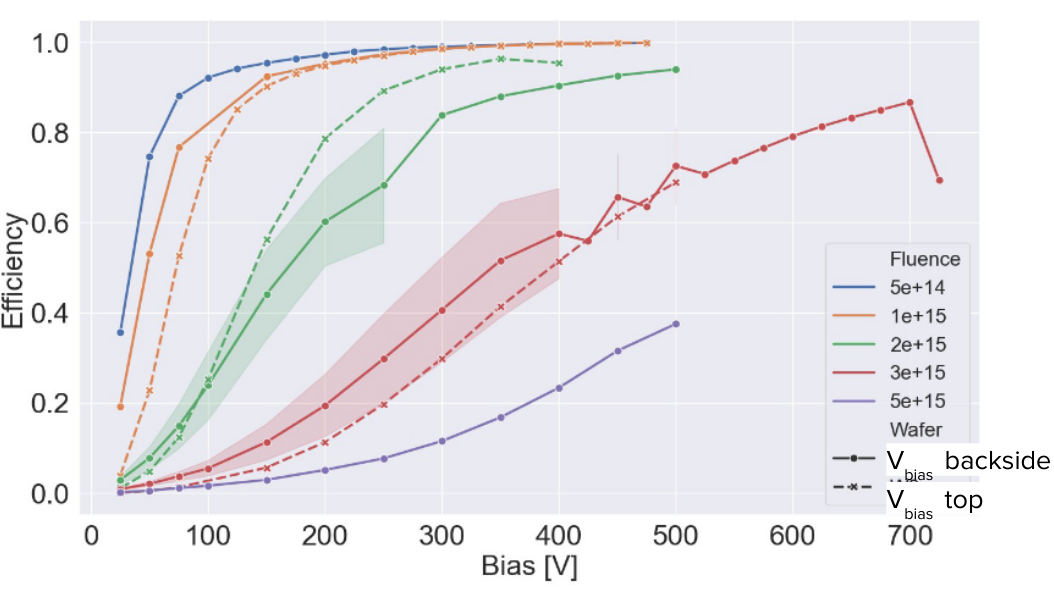}
    \caption{Efficiency of the RD50-MPW5 for varying thresholds (left) and bias voltages (right) \cite{vciproceedings}.}
         \vspace{-0.7cm}
    \label{fig:eff_vs_bias_thr}
\end{figure}

The efficiency of the chip was also measured against threshold and bias voltage, as shown in \cref{fig:eff_vs_bias_thr}. An efficiency of more than 99\% was reached for fluences between $\Phi = 1 \times 10^{15}~\mathrm{n}_{\mathrm{eq}}/\mathrm{cm}^2$ and $\Phi = 2 \times 10^{15}~\mathrm{n}_{\mathrm{eq}}/\mathrm{cm}^2$ for acceptable thresholds and bias voltages. Here backside biased sensors performed mostly better than topside biased sensors. An acceptable fake hit rate was found for sensors irradiated up to $\Phi = 3 \times 10^{15}~\mathrm{n}_{\mathrm{eq}}/\mathrm{cm}^2$. Sensors were affected by noise from $\Phi = 5 \times 10^{15}~\mathrm{n}_{\mathrm{eq}}/\mathrm{cm}^2$.

\vspace{-0.3cm}
\section{3D Image of the Sensor}
\label{sec:tpa}
A 3D image was also made of the RD50-MPW4 sensor using the Two Photon Absorption-Transient Current Technique \cite{tpampw4}. The TPA-TCT employs a femtosecond pulsed laser in the near infrared spectrum. Such lasers are available in only a limited number of high energy physics institutes, among them Nikhef and CERN, but also JSI Ljubljana, IFCA, Lancaster, Oxford and Manchester. This measurement was made at the CERN Solid State Detector Lab using backside illumination with a 1550 nm, 400 fs pulse laser. For this purpose, a non-backside processed sensor was used to avoid reflections. A hexapod stage was used for moving the sensor in XY and Z, and pulses with 0.25~nJ were injected at a rate of 1 kHz. The sensor was operated at a threshold of 940 mV and a bias voltage of -90~V. 
\begin{figure}
    \centering
    \includegraphics[width=0.37\linewidth]{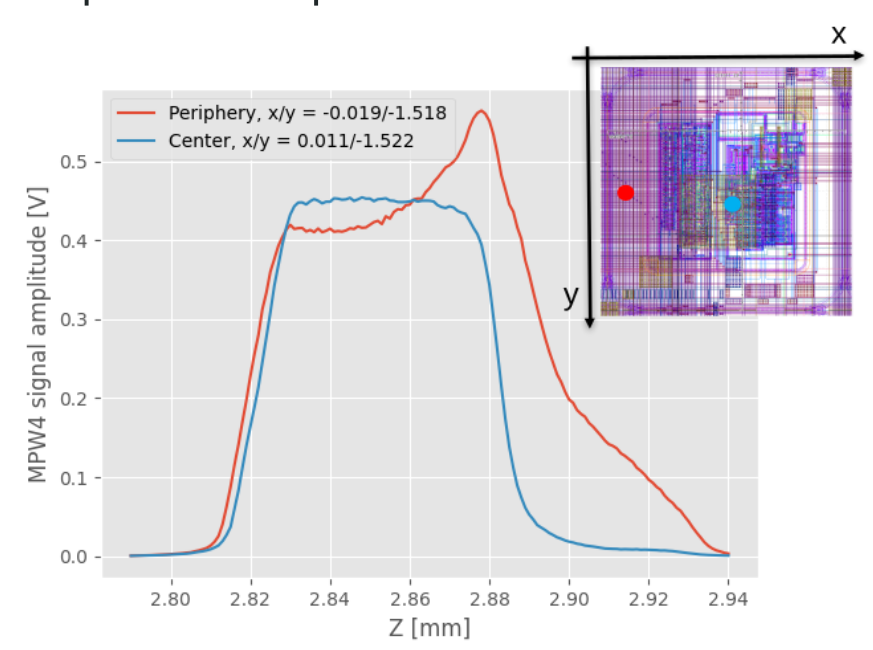}    
    \includegraphics[width=0.32\linewidth]{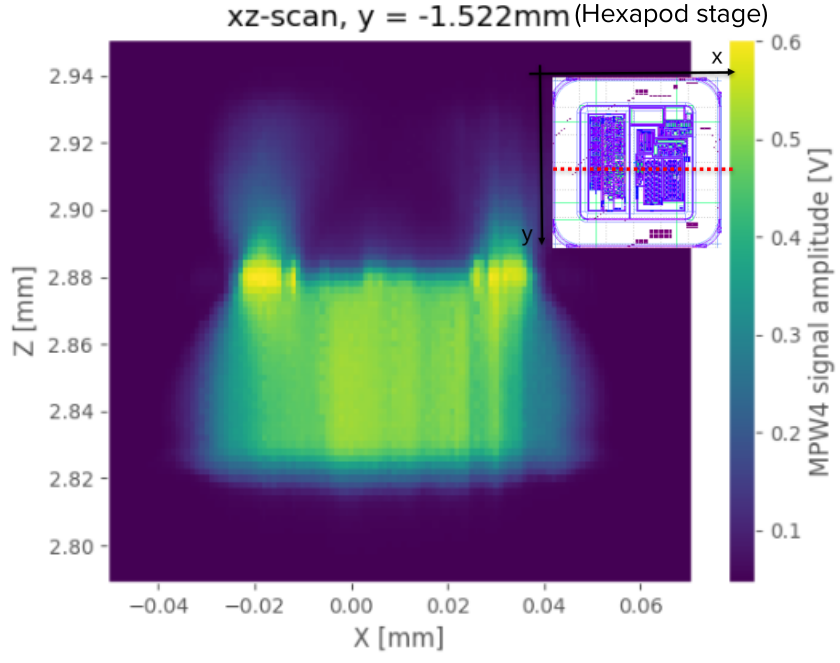}
    \caption{Scan from backside to topside of the sensor for pixel (31,33) from 0 to \SI{264}{\micro\meter}, with x scanned from 0 to \SI{62}{\micro\meter}. Y is kept at \SI{85}{\micro\meter}. The electronics is clearly visible especially at the periphery of the sensor, and the depletion depth of \SI{226}{\micro\meter} at a bias voltage of -90~V.}
     \vspace{-0.7cm}
    \label{fig:zscan}
\end{figure}

Scanning Z from the backside to the top from 0 to \SI{264}{\micro\meter}, and X from 0 to \SI{62}{\micro\meter}, keeping Y at \SI{85}{\micro\meter}, one can clearly see the reflection at the metals (bump in red on the left in \cref{fig:zscan}) and the depletion depth of \SI{226}{\micro\meter} at a bias voltage of -90~V (yellow signal region on the right in \cref{fig:zscan}). The measurement was done for pixel (33,31), and the Z-scale is corrected for refraction in silicon \cite{tpameasurements}. 

\begin{figure}
    \centering
    \includegraphics[width=0.25\linewidth]{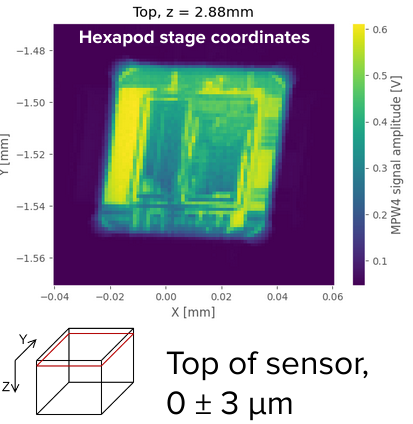}
        \includegraphics[width=0.27\linewidth]{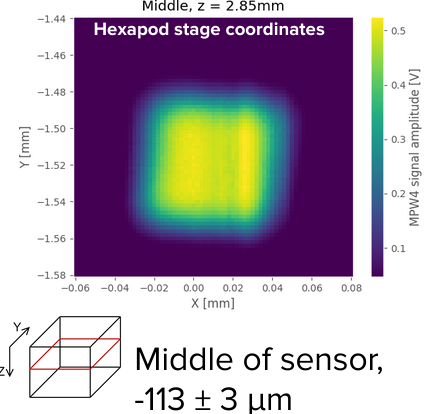}
    \includegraphics[width=0.27\linewidth]{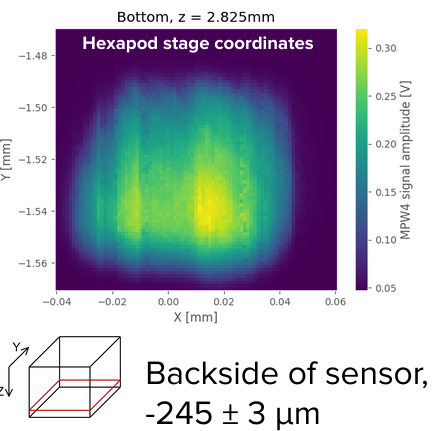}
    \caption{XY scan over entire pixel (31, 33) at three different depths in the sensor. Electronics is clearly visible at the top, and the edge of the depletion region is clearly visible at the backside of the sensor.}
            \vspace{-0.7cm}
    \label{fig:xyscan}
\end{figure}

The signal amplitude was also measured at different pixel depths, where the metal clearly affects the signal response. The electronics is visible at the top (see \cref{fig:xyscan}, left) and close to the backside of the sensor at -\SI{245}{\micro\meter} the edge of the depletion region is reached (see \cref{fig:xyscan}, right). 

\vspace{-0.3cm}

\section{Summary and Outlook}
\label{sec:conclusion}
In summary, the RD50-MPW are a series of monolithic High Voltage CMOS pixel sensors with breakdown voltages over 600~V. They can be fully depleted and have a high radiation tolerance with high voltage and backside processing. The chips were found to have more than 99\% efficiency for fluences up to $\Phi = 2 \times 10^{15}~\mathrm{n}_{\mathrm{eq}}/\mathrm{cm}^2$. A 3D image of the sensor was also obtained using the TPA-TCT technique. R\&D with the LFoundry 150~nm process will continue within the DRD3 collaboration with the RadPix sensor towards the LHCb Mighty Tracker.
\vspace{-0.3cm}
\section{Acknowledgements}
This work has been partly performed in the framework of the CERN-RD50 collaboration.








\vspace{-0.36cm}
\bibliographystyle{model1-num-names}

\begin{thebibliography}{99}







\bibitem{rd50hvcmos} CERN-RD50 collaboration, "Development of high voltage-CMOS sensors within the CERN-RD50
collaboration", \textit{Nucl. Instrum. Meth. A} \textbf{1034} (2022) 166826, doi:\href{https://doi.org/10.1016/j.nima.2022.166826}{10.1016/j.nima.2022.166826}.

\bibitem{mpw2timeres} J. Debevc \textit{et al.}, "Measurements of time resolution of the RD50-MPW2 DMAPS prototype using TCT and 90Sr", \textit{JINST} \textbf{19} (2024) P05068, doi:\href{https://doi.org/10.1088/1748-0221/19/05/P05068}{10.1088/1748-0221/19/05/P05068}.
\bibitem{dmaps180} J. Dingfelder \textit{et al.}, "Progress in DMAPS developments and first tests of the Monopix2 chips in 150 nm LFoundry and 180 nm TowerJazz technology", \textit{Nucl. Instrum. Meth. A} \textbf{1034} (2022) 166747, doi:\href{https://doi.org/10.1016/j.nima.2022.166747}{10.1016/j.nima.2022.166747}.
\bibitem{guardrings} S. Zhang \textit{et al.}, "Breakdown performance of guard ring designs for pixel detectors in 150 nm CMOS technology", \textit{Nucl. Instrum. Meth. A} \textbf{1063} (2024) 169287, doi:\href{https://doi.org/10.1016/j.nima.2024.169287}{10.1016/j.nima.2024.169287}.

\bibitem{mpw4} E. Vilella \textit{et al.}, "RD50-MPW4: a thin backside-biased High Voltage CMOS pixel chip for high radiation tolerance", \textit{JINST} \textbf{20} (2025) C03044, doi:\href{https://doi.org/10.1088/1748-0221/20/03/C03044}{10.1088/1748-0221/20/03/C03044}
\bibitem{mpw4presentationchenfan} Chenfan Zhang, " Summary of the RD50-MPW Activities and last Results ", \href{https://indico.cern.ch/event/1492202/contributions/6287606/}{AIDAinnova Final Annual Meeting}, Prague, 2025.


\bibitem{cmosthesis} L. Meng, "Development of CMOS Sensors for High-Luminosity ATLAS Detectors", Ph.D. Thesis, University of Liverpool (2018) [\href{https://cds.cern.ch/record/2637873}{CERN-THESIS-2018-153}].
\bibitem{etcthvcmos} M. Franks \textit{et al.}, "E-TCT characterization of a thinned, backside biased, irradiated HV-CMOS pixel test structure, \textit{Nucl. Instrum. Meth. A} \textbf{991} (2021) 164949, doi:\href{https://doi.org/10.1016/j.nima.2020.164949}{10.1016/j.nima.2020.164949}.
\bibitem{ccecmos} I. Mandić \textit{et al.}, "Charge collection properties of irradiated depleted CMOS pixel test structures", \textit{Nucl. Instrum. Meth. A} \textbf{903} (2018) 126, doi:\href{https://doi.org/10.1016/j.nima.2018.06.062}{10.1016/j.nima.2018.06.062}.
\bibitem{vciproceedings} B. Pilsl \textit{et al.}, "Enhancing radiation hardness and granularity in HV-CMOS: The RD50-MPW4 sensor", \textit{Nucl. Instrum. Meth. A} \textbf{1080} (2025), 170752, doi:\href{https://doi.org/10.1016/j.nima.2025.170752}{10.1016/j.nima.2025.170752}
\bibitem{mpwseries} C. Zhang \textit{et al.}, "RD50-MPW: a series of monolithic High Voltage CMOS pixel chips with high granularity and towards high radiation tolerance", \textit{JINST} \textbf{19} (2024) C04059, doi:\href{https://doi.org/10.1088/1748-0221/19/04/C04059}{10.1088/1748-0221/19/04/C04059}.
\bibitem{mpw4characterization} B. Pilsl \textit{et al.}, "Characterization of the RD50-MPW4 HV-CMOS pixel sensor", \textit{Nucl. Instrum. Meth. A} \textbf{1069} (2024) 169839, doi: \href{https://doi.org/10.1016/j.nima.2024.169839}{10.1016/j.nima.2024.169839}.


\bibitem{caribou} Y. Otarid \textit{et al.}, "Caribou -- A versatile data acquisition system for silicon pixel detector prototyping", \textit{JINST} \textbf{20} (2025) C07043, doi:\href{ https://doi.org/10.1088/1748-0221/20/07/C07043}{10.1088/1748-0221/20/07/C07043}
.

\bibitem{adenium} Y. Liu \textit{et al.}, "ADENIUM -- A demonstrator for a next-generation beam telescope at DESY", \textit{JINST} \textbf{18} (2023) P06025, doi:\href{https://doi.org/10.1088/1748-0221/18/06/P06025}{10.1088/1748-0221/18/06/P06025}.
\bibitem{telepix} H. Augustin \textit{et al.}, "TelePix -- A fast region of interest trigger and timing layer for the EUDET Telescopes", \textit{Nucl. Instrum. Meth. A} \textbf{1048} (2023) 167947, doi:\href{https://doi.org/10.1016/j.nima.2022.167947}{10.1016/j.nima.2022.167947}.
\bibitem{eudaq2} Y. Liu \textit{et al.}, "EUDAQ2 -- A flexible data acquisition software framework for common test beams", \textit{JINST} \textbf{14} 2019 P10033 doi:\href{https://doi.org/10.1088/1748-0221/14/10/P10033}{10.1088/1748-0221/14/10/P10033}.
\bibitem{corry} D. Dannheim \textit{et al.}, "Corryvreckan: A Modular 4D Track Reconstruction and Analysis Software for Test Beam Data", \textit{JINST} \textbf{16} (2021) P03008, doi:\href{https://doi.org/10.1088/1748-0221/16/03/P03008}{10.1088/1748-0221/16/03/P03008}.
\bibitem{resolution}  T. Alexopoulos \textit{et al.}, "Examining the geometric mean method for the extraction of spatial resolution", \textit{JINST} \textbf{9} (2014) P01003, doi:\href{http://doi.org/10.1088/1748-0221/9/01/P01003}{10.1088/1748-0221/9/01/P01003}.
\bibitem{tpampw4} F.R. Palomo \textit{et al.}, "TPA-TCT Analysis of the RD50-MPW4 Monolithic Pixel Particle Detector", (2026) doi:\href{https://doi.org/10.48550/arXiv.2605.11961}{10.48550/arXiv.2605.11961}.
\bibitem{tpameasurements} M. Wiehe \textit{et al.}, "Development of a Tabletop Setup for the Transient Current Technique Using Two-Photon Absorption in Silicon Particle Detectors", \textit{IEEE Trans. Nucl. Sci.} \textbf{68} (2021) 2, doi:\href{https://doi.org/10.1109/TNS.2020.3044489}{10.1109/TNS.2020.3044489}.


\end{thebibliography}

\end{document}